\documentclass[a4paper,fleqn]{mnras}

\usepackage[T1]{fontenc}
\usepackage{ae,aecompl}

\usepackage{graphicx}	
\usepackage{amsmath}	
\usepackage{amssymb}	
\usepackage{mathtools, cuted}
\usepackage{lipsum}

\title[Dynamics of Magnetic Flux Tubes]{Dynamics of Magnetic Flux Tubes in an Advective Flow around a Black Hole }

\author[Arnab Deb, Kinsuk Giri and Sandip K. Chakrabarti]
{Arnab Deb\thanks{arnab12a@bose.res.in}$^{1}$, Kinsuk Giri\thanks{kinsuk@nitttrkol.ac.in}$^{2}$, Sandip K. Chakrabarti \thanks{chakraba@bose.res.in}$^{1,3}$\\
$^{1}$ S. N. Bose National Centre for Basic Sciences, Block -JD,  Sector -3, Salt Lake, Kolkata 700106, India\\
$^{2}$ National Institute of Technical Teachers Training and Research, Block-FC, Sector - 3, Salt Lake, Kolkata-700106, India \\
$^{3}$ Indian Centre for Space Physics, Chalantika 43, Garia Station Road, Kolkata-700084, India}

\pubyear{2016}

\begin{document}
\label{firstpage}
\pagerange{\pageref{firstpage}--\pageref{lastpage}}
\maketitle

\begin{abstract}
Entangled magnetic fields entering into an accretion flow would very soon be stretched into 
a dominant toroidal component due to strong differentially rotating motion inside the accretion disc.
This is particularly true for weakly viscous, low angular momentum transonic or advective discs.
We study the trajectories of toroidal flux tubes inside a geometrically thick 
flow which undergoes a centrifugal force supported shock. We also study effects of these 
flux tubes on the dynamics of the inflow and the outflow. We use a finite difference method (Total 
Variation Diminishing) for this purpose and specifically focussed on whether these flux tubes 
significantly affect the properties of the outflows such as its collimation and the rate. 
It is seen that depending upon the cross-sectional radius of the flux tubes which control 
the drag force, these field lines may move towards the central object or oscillate 
vertically before eventually escaping out of the funnel wall (pressure zero surfaces) along the vertical direction.
A comparison of results obtained with and without flux tubes show these flux tubes could play pivotal role 
in collimation and acceleration of jets and outflows.  
\end{abstract}



\section{Introduction}

A clear understanding of formation, acceleration, and collimation of radio jets
has been an ever eluding problem for the astrophysicists. Early theoretical approach to
collimate jets came from the study of geometrically thick accretion discs.
In such a disc, a vortex structure is created due to the strong centrifugal 
barrier near the axis (Lynden-Bell 1978) and it is assumed that some disc matter is 
continuously accelerated due to super-Eddington luminous radiation in this opening. 
The collimation and acceleration of this ejected matter 
is aided by hydrodynamic (Fukue 1982, Chakrabarti 1986) or
hydro-magnetic processes (Blandford \& Payne 1982, Chakrabarti \& Bhaskaran 1992).
Camenzind (1989), and Heyvaerts \& Norman (1989) Lovelace (1976)
show that the magnetic field contributes to the collimation of jets as well. The self-similar magnetic-field 
structure inside a cold, partially ionized disc was determined by K\"onigl (1989)
where it was shown that for a certain set of disc parameters
Blandford \& Payne (1982) type self-similar winds can be produced which can achieve 
super-Alfv\'enic velocity at a finite distance from the disc surface. Magnetic fields 
are believed to be brought inside the the disc from the companion 
and its environment and after amplification, compression and advection they are finally 
eliminated from the disc due to magnetic buoyancy. 
A large body of literature is present that explores many aspects of 
the effects of buoyancy and shear amplification on these magnetic flux in the paradigm of thin accretion 
disc.(e.g., Eardley \& Lightman 1975, Galeev et al. 1979, Coroniti 1981, 
Shibata etal. 1990, Chakrabarti et al. 1994). 
If the buoyancy timescale is larger than the shear amplification time scale, 
flux tubes can be amplified exponentially until they reach the equipartition 
value, i.e., where magnetic pressure matches with local gas pressure.
Detailed study of the dynamics of the flux tubes in the context of thin accretion discs 
has been done by Sakimoto \& Coroniti (1989). Study of behaviour of toroidal magnetic flux tubes in the 
backdrop of thick accretion discs has been done by Chakrabarti \& D'Silva (1994a, hereafter CD94(I))
and Chakrabarti \& D'Silva (1994b, hereafter, CD94(II)) where they showed that when a flux tube is injected into the thick 
disc outer boundary it will emerge in the 'chimney' (the funnel like opening in the inner part of the disc) 
depending on the various flows and field parameters, such as, initial position of release, 
cross-sectional radius of the flux tubes, angular momentum distribution,
etc. In these latter studies, the disc thermodynamic parameters were time independent while flux tube 
is allowed to be advected in with a suitably chosen radial flow. So far, no study of the flux 
tube behaviour and its possible effects on the flow dynamics and on the jet formation in the context of a self-consistent, 
time dependent, geometrically thick transonic flow has been performed. 

In the present paper, we study how and where the flux tubes emerge from the accretion disc 
and whether it plays any role in collimation of jets and outflows. We investigate the circumstances under which 
the flux tubes released at the outer boundary would appear in the chimney. 
Since we have only assumed the inviscid flow which conserves the angular momentum, 
no Coriolis force will act on the flux tubes in local rotating frame (CD94(I)). For this reason, the 
buoyant flux tubes will rise along direction of the pressure gradient force, bringing only a fraction of flux 
tubes into the chimney. In presence of viscosity, Coriolis force will also play a major role in 
shaping the dynamics of the flux tubes (CD94(II)). In case of Sun, since the magnetic fields 
are anchored, a sustained magnetic activity can be seen as flux tubes could partially emerge
through Parker instability (Parker, 1955). CD94(II) have 
shown that under certain circumstances, magnetic fields can also be anchored inside the thick disc as 
they tend to oscillate about the equipotential surface so as to produce a stable corona and such instability may arise.
However, our presently assumed axisymmetric flow properties preclude this study.

In order to study the dynamics of the flux tubes we have considered an advective disc which 
can be obtained from time dependent solutions of Navier-Stokes equation. A pseudo-Newtonian 
potential as prescribed by Paczy\'nsky \& Wiita (1980) is used to mimic the space time 
external to the black hole. The flux tubes are assumed to be axisymmetric about the rotation
axis of the disc and have only azimuthal component of the field (CD94(I)). In the next three Sections 
we present equations governing the disc, equation of motion of the flux tubes, and simulation procedures. In
\S 5, we present numerical results and in the last Section we summarize our results and discuss the implications.        
\section{Model Equation}
 For the purpose of the present study, we consider a two dimensional axisymmetric accretion flow around a stellar mass schwarzschild black hole. As prescribed by Paczy\'nski \& Wiita (1980), here we use a pseudo Newtonian potential which is found to be very efficient in mimicking the effects of general relativity (see, Ryu, Chakrabarti \& Molteni (1997, hereafter, RCM97), Giri \& Chakrabarti (2010, hereafter, GC10). We use cylindrical polar coordinates  ($r$, $\phi$ and $z$) for our calculations. We choose the mass of the black hole ($M_{BH}$), velocity of the light ($c$), and the Schwarzschild radius $r_g=2GM_{BH}/c^2$  as the units of the mass, velocity and distance respectively.

The hydrodynamic equations of an inviscid flow comprising of the mass, momentum, and energy conservation equations 
in a compact form using dimensionless units are presented in Ryu et al. (1995), Molteni, Ryu \& Chakrabarti
(1996, hereafter, MRC96), Giri \& Chakrabarti (2010, hereafter, GC10) and Giri (2014) in great detail. The hydrodynamic equations can be obtained 
from Navier-Stokes equation by ignoring viscous stress tensor. The hydrodynamic equations for inviscid flow are given as,

\begin{equation}
  \frac{\upartial \rho}{\upartial t}+\nabla . (\rho \mathbf{v})=0
\end{equation} 
 
\begin{equation}
  \rho\frac{\upartial \mathbf{v}}{\upartial t}+\rho(\mathbf{v}. \nabla ).\mathbf{v})= -\nabla p+\mathbf{f_{grav}}+\mathbf{f_{mag}}
\end{equation} 

\begin{equation}
  \frac{\upartial E}{\upartial t}+\nabla.\bigg[\left(H+\frac{1}{2}v^2\right).\rho \mathbf{v}\bigg]=0
\end{equation}
 
Since we are interested in inviscid magnetized flow, here we have included force due to the presence of magnetic field. $H$ is the
enthalpy of the system, $\mathbf{f_{grav}}$ and $\mathbf{f_{mag}}$ are the gravitational force and the force due to magnetic field respectively.
Equations 1, 2, and 3 can be written in terms of conservative variables in a compact form,

\begin{equation}
{\upartial{\mathbfss q}\over\upartial t}+{1\over r}{\upartial\left(r
{\mathbfss F}_1\right)\over\upartial r}+{\upartial{\mathbfss F}_2\over\upartial r}
+{\upartial{\mathbfss G}\over\upartial z} = {\mathbfss S},
\end{equation} 

where the state vector is

\begin{equation}
{\mathbfss q} = 
                          \begin{pmatrix}
                          \rho\\
                          \rho v_r\\
                          \rho v_{\theta}\\
                          \rho v_z\\
                           E
                           \end{pmatrix},
\end{equation}
the flux functions are
\begin{equation}
{\mathbfss F}_1 =            
                         \begin{pmatrix}
                         \rho v_r\\
                         \rho v_r^2\\
                         \rho v_{\theta}v_r\\
                         \rho v_z v_r\\
                         (E+p)v_r\\ 
                          \end{pmatrix} \qquad,
{\mathbfss F}_2 = 
                         \begin{pmatrix}
                          0\\
                          p\\
                          0\\
                          0\\
                          0\\
                          \end{pmatrix} \qquad,
{\mathbfss G} =  
                         \begin{pmatrix}
                         \rho v_z\\
                         \rho v_r v_z\\
                         \rho v_{\theta} v_z\\
                         \rho v_z^2+p\\
                         (E+p)v_z\\
                         \end{pmatrix} ,
\end{equation} \\
and the source function is

\begin{equation}
{\mathbfss S} =   
                \begin{pmatrix}
                   0\\
                ~~~\\
                {\rho v_{\theta}^2\over r}
                -{{\rho r}\over 2 \left(\sqrt{r^2+z^2}-1\right)^2\sqrt{r^2+z^2}}-f_{mag,r}\\
                ~~~\\
                ~~~\\
                -{\rho v_r v_{\theta}\over r}\\
                ~~~\\
                -{\rho z\over2\left(\sqrt{r^2+z^2}-1\right)^2\sqrt{r^2+z^2}}+f_{mag,z}\\
                ~~~\\
                ~~~\\
                -{\rho \left(rv_r+zv_z\right)\over
                2\left(\sqrt{r^2+z^2}-1\right)^2\sqrt{r^2+z^2}}\\
                \end{pmatrix} .
\end{equation}

Here, the energy density (without potential energy) is given as(GC10, MRC96), $$E=p/(\gamma-1)+\rho(v_r
^2+v_{\theta}^2+v_z^2)/2,$$ 
where  $\rho$ is the mass density, $\gamma$ is the adiabatic index, $p$ is the pressure, 
$v_r$, $v_\theta$ and $v_z$ are the radial, azimuthal and vertical component of velocity 
respectively. Now, since we have considered an inviscid flow, this implies that the specific angular momentum ($\lambda$) is conserved i.e., $$\frac{d\lambda}{dt}=0$$.
In a black hole accretion, matter with specific angular momentum up to the marginally stable value ($1.83$ in units of $2GM/c$) does not require any viscosity to accrete. These are sub-Keplerian flows which work as the Comptonizing component in a black hole accretion. The Keplerian component will still be present on the equatorial plane if viscosity is high enough (as shown in Giri et al. 2013).  However,  in the present paper we do not consider any Keplerian disk. Here we have only sub-Keplerian matter advected towards the black hole. Thus the accretion rate is only for sub-Keplerian halo that is advected towards the compact object, not of the conventional Keplerian disk.
In the context of the present paper the source function is changed since we are interested 
in magnetized inviscid flow but magnetic field is not permeated everywhere in the flow. 
It is only restricted to the flux tubes whose dynamics we are studying. Thus the Lorentz 
force
is operative only due to them. The source function $\mathbfss S$ bears two terms $f_{mag,r}$ 
and $f_{mag,z}$ which are $r$ and $z$ components of the Lorentz force due to the presence 
of toroidal magnetic field. The expression for the Lorentz force is given as, 
\begin{equation}
\mathbf{f}_{mag} = \frac{\rho}{(m_{electron}+m_{proton})}\frac{e}{c}(\mathbf{v}\times\mathbf{B}).
\end{equation}  

The general form of the equations of the flow in an inertial reference frame (Batchelor 1967) is given by,
\begin{equation}
\rho [{\partial {\bf v} \over \partial t} + {{\bf v} . {\nabla {\bf v}}}] = 
- { \nabla p} + {\bf {F_b}}  + {\nabla . {\bf {\tau}}} ,
\end{equation}
where, ${\bf v}$ is the flow velocity, ${\bf \tau}$ is the viscous stress tensor, and ${\bf {F_b}}$  
represents body forces (per unit volume) acting on the fluid and ${\nabla}$ is the Del operator.
Here, the body forces consist of only gravity forces and electromagnetic 
forces (Lorentz force). We are considering only magnetized inviscid flow in the present paper and therefore, 
detailed discussion on viscous stress tensor and its influence on the flow solution is out of the scope 
of this paper. We also assume a polytropic equation of state
for the accreting (or outflowing) matter, $p = K\rho^{\gamma}$ , where, $p$ and
$\rho$ are the pressure and the fluid density, respectively and $\gamma$ is
the adiabatic index which is connected to polytropic index($n$) by the expression $\gamma=1+\frac{1}{n}$

As discussed before, the Pseudo-Newtonian gravitational potential for a point mass sitting at
origin in cylindrical coordinate system described as Paczy\'nsky \& Wiita (1980),
\begin{equation}
\phi(r,z) = -{GM_{BH}\over(R-r_g)}, 
\end{equation}
where, $R=\sqrt{r^2+z^2}$, is used for the purpose of this paper. 

\section{Equation of Motion of the Flux tubes}

 Parker (1955) in his pioneering work showed how magnetic buoyancy causes internally generated flux tubes to come out causing different solar phenomena. On the 
contrary, in case of advective accretion disc around a black hole the picture is somewhat 
different. Unlike the sun, in case of black hole the flux tubes are not produced within 
the disc, instead they are brought along with the matter accreted from the companion and
also effect of differential rotation in the disk
is much more pronounce. In the past few years, a significant amount
of work has been done regarding the stability, dynamics of magnetic flux tubes and their 
different aspects and effects on astrophysical jets (e.g., Shibata \& Uchida (1985), Shibata \& Uchida (1986), Ferriz-Mas et al. (1989), 
Moreno-Insertis et al. (1992), You et al. (2005), Longcope \& Klapper (1997), Blackman (1996), Fendt \& Camenzind (1996), CD94(I) \& CD94(II), 
Choudhuri \& Gilman (1987) etc.). 

We consider an azimuthally symmetric flux ring and we use thin flux tube approximation 
which lets us assume that the variation of different physical quantities inside the tube
is negligible. The approximation is valid if the cross-sectional radius of the flux ring
is smaller compared to the local pressure scale height of the disc.
Due to differential 
rotation (DR)  and turbulent eddies (TE) two important things happen  inside a thick disk.  
DR would stretch isolated blob of magnetic fields into mostly azimuthal, oppositely directed 
fields which the TE's may push and reconnect and increase the number of blobs. This is not 
applicable for axisymmetric field lines. However, a single thick toroidal field lines may be 
split into several axisymmetric fillaments in presence of shear and the number of axisymmetric 
flux tubes may increase. In that case, the number of the flux tubes will not be conserved. 
However we do not assume tearing up of a field into many as that would involve application of 
detailed turbulent theory and since the turbulence cells are much 
smaller than the chosen gridsize as well as the cross sectional radius of the flux tube, 
which is beyond the scope of this paper. The magnetic flux tubes of random size brought 
in by advection can be sheared which will be discussed elsewhere. 
 
The equations of motion for the thin flux tubes have been written by several authors in
the context of solar physics as well as thick disc around a black hole (Choudhuri \& Gilman 1987, CD94(I)). Here, following CD94(I), 
the equations of motion for thin axisymmetric flux tube are given as,

\begin{equation}
\begin{aligned}
\ddot{\xi}-\xi\dot{\theta}+\frac{X}{(1+X)}[-\xi{\dot{\phi}}^2\sin^2\theta-2\xi\omega\dot{\phi}\sin^2\theta]= \\
\frac{X}{(1+X)}\bigg\{\frac{M}{X}[g-\xi{\omega}^2\sin^2\theta]-T_{ens}\sin\theta-\frac{D_{r}}{\pi{\sigma}^2\rho_{e}}\bigg\}, 
\end{aligned}
\end{equation} 

\begin{equation}
\begin{aligned}
\xi\ddot{\theta}+2\dot{\xi}\dot{\theta}+\frac{X}{(1+X)}[-\xi{\dot{\phi}}^2\sin\theta\cos\theta-2\xi\omega\dot{\phi}\sin\theta\cos\theta]= \\
-\frac{X}{(1+X)}\bigg\{\frac{M}{X}\xi{\omega}^2\sin\theta\cos\theta+T_{ens}\cos\theta+\frac{D_{\theta}}{\pi{\sigma}^2\rho_{e}}\bigg\}, 
\end{aligned}
\end{equation} 

\begin{equation}
\begin{aligned}
\xi\sin\theta\ddot{\phi}+2\dot{\xi}\sin\theta(\dot{\phi}+\omega)+2\xi\cos\theta\dot{\theta}(\dot{\phi}+\omega)+
\xi\sin \theta \\
 \bigg\{\dot{\xi}\frac{\partial\omega}{\partial r}+\dot{\theta}\frac{\partial \omega}{\partial \theta}\bigg\} = 0,
\end{aligned}
\end{equation}
where, $(\xi, \theta, \phi)$ is the position of a point inside a 
flux ring having magnetic field $B$. Here, $\xi$ is measure of radial distance in the 
unit of Scwarzschild radius ($r_g$). 
Here, $X$ is defined as $X = m_{i}/m_{e}$ where, $m_i=\rho_i\pi\sigma^2\cdot2\pi r
\sin\theta$ is the mass inside the flux tube of radius of cross section (in the meridional plane) 
$\sigma$ and $m_e=\rho_e\pi\sigma^2\cdot2\pi r\sin\theta$ is the mass of the external 
flow
displaced by the flux tube. The flux $\psi=B\pi\sigma^2$ through the ring remains 
constant. 
The flux tube experiences buoyancy and the buoyancy factor is given by, 
$$ M= \frac{\rho_e-\rho_i}{\rho_e}=\frac{m_e-m_i}{m_e}$$ 
where $\rho_e$ and $\rho_i$ represent external and internal densities respectively. 
The effective acceleration due to gravity is,
\begin{equation}
\mathbfit{g}_{eff}=(g-\xi\omega^2\sin^2\theta)\hat{\xi}-\xi\omega^2\sin\theta\cos
\theta\hat{\theta}
\end{equation}    
where, $g$ is given as $g= 1/2(\xi-1)^2$ in Schwarzschild unit.
The drag force per unit length is given as,
\begin{equation}
\mathbfit{D}=-0.5C_D\rho_e\sigma\vert(\dot{\xi}-v)\hat{\xi}+\xi\dot{\theta}\hat
{\theta}\vert{(\dot{\xi}-v)\hat{\xi}+\xi\dot{\theta}\hat{\theta}}
\end{equation}
where, $C_D = 0.4$ is a dimensionless coefficient that has constant value of $0.4$ for 
high Reynold's number, (Goldstein, 1938).
It is often said that a magnetic field line is like a rubber band. Just as a stretched 
band returns to its original size when released, a closed magnetic flux tube also has a 
tension which is often the most important force.  The tension force is given by
$$T_{ens}=\frac{4\pi M_0 T_e(\xi_0)}{\mu_e A(1-M_{0})
\xi_0\sin\theta_0}$$ is a dimensionless measure of the magnetic tension (CD94(I)) 
where, $T_e(\xi_0)$ is the initial temperature of the external fluid, ($\xi_0,\theta_0$) 
is the initial position of the flux tube, $A$ is area increment factor given as 
$A=(\sigma/\sigma_0)^2$ where $\sigma_0$ is the initial cross sectional radius and $\sigma$ 
is the instantaneous radius, and  $M_0$ is the initial buoyancy factor, which is 
calculated to be $M_0=B^2 / 8\pi p_{g,e}$ where, $p_{g,e}$ is external gas pressure.
 As $A(t) = \sigma^2(t)/\sigma^2_0$ , this gives the evolution of the flux 
tubes over the course of the simulation. \textbf{Explicit form of this area 
expansion factor is given as (CD94(I)),} 
$$A = \bigg(\frac{T_e(\xi_0,\theta_0)}{T_e(\xi,\theta)}
\bigg)^3\bigg(\frac{\xi_0\sin\theta_0}{\xi\sin\theta}\bigg)\bigg(\frac{1-M_0}{1-M}\bigg)$$
where, $T_e(\xi,\theta)$ is directly coming from our simulation at each instant of time.

\section{Simulation Procedures}

In order to proceed further, we formulate first how to calculate the 
magnetic buoyancy factor ($M$) and the area increment factor ($A$). 
The magnetic buoyancy $M$ depends on the process of energy transfer between
the disc and flux tube and entropy distribution. Here, we have considered the 
flux tubes to be moving adiabatically, i.e., there is no heat exchange between
the flux tubes and its surroundings. The entropy inside remains constant throughout 
the path 
traversed within the disc.Using the fact that the flux tube is in pressure equilibrium
with the surroundings we have,
\begin{equation}
  p_{r,i}+p_{g,i}+\frac{B^2}{8\pi}=p_{r,e}+p_{g,e},
\end{equation}
where, $p_{r,i}$ and $p_{r,e}$ are radiation pressure for internal and external fluid 
and $p_{g,i}$ and $p_{g,e}$ are gas pressure for internal and external fluid. The ratio
of gas pressure ($p_g$) to total pressure ($p_r+p_g$) is denoted by a constant
 $\beta$. 
From eq.(12), using the assumption that the flux tube is in thermal equilibrium with the
surrounding just prior to its release, we get,
\begin{equation}
\frac{B}{\rho_{i}\xi\sin\theta}= constant
\end{equation} 
Since the flux tube moves adiabatically, we have,
\begin{eqnarray}
\frac{p_i}{p_{i0}}=\bigg(\frac{\rho_i}{\rho_{i0}}\bigg)^{\gamma} \\
\frac{p_e}{p_{e0}}=\bigg(\frac{\rho_e}{\rho_{e0}}\bigg)^{\gamma} \\
\frac{T_i}{T_{i0}}=\bigg(\frac{\rho_i}{\rho_{i0}}\bigg)^{\gamma-1} \\
\frac{T_e}{T_{e0}}=\bigg(\frac{\rho_e}{\rho_{e0}}\bigg)^{\gamma-1} 
\end{eqnarray}
where, $p$ and $T$ represents pressure and temperature respectively and the subscripts
$i,~e,~ \&~ 0$ represents flow inside the tube, external flow, and initial values 
of the physical quantities respectively. After rearranging and then dividing eq.12
by $p_{g,e}$ we get,
\begin{equation}
\frac{1-\beta}{\beta}\bigg(1-\frac{p_{r,i}}{p_{r,e}}\bigg)+1-\frac{p_{g,i}}{p_{g,e}}=
\frac{B^2}{8\pi p_{g,e}}
\end{equation}
putting $p_r=\frac{1}{3}aT^4$ and using the fact that $T_{i0}=T_{e0}$ in Eq. 18 we get,
\begin{equation}
\frac{1-\beta}{\beta}\bigg(1-\frac{(T_{i}/T_{i0})^4}{(T_{e}/T_{e0})^4}\bigg)+1-\frac
{(p_{g,i}/p_{i0})}{(p_{g,e}/p_{e0})}\frac{p_{i0}}{p_{e0}} = \frac{B^2}{8\pi p_{g,e}} 
\end{equation}
Using eq. 13-17 into Eq.18 we get an expression for $\rho_i/\rho_e$,
\begin{equation}
k_1\bigg(\frac{\rho_i}{\rho_e}\bigg)^{4/3} + k_2\bigg(\frac{\rho_i}{\rho_e}\bigg)^2 
-1 = 0, 
\end{equation}
where,
$$
k_1=\frac{(1-\beta_e M_0)}{(1-M_0)^{4/3}},$$ $$k_2=\beta_e \frac{M_0}{(1-M_0)^2
}\bigg(\frac{T_e}{T_{e,0}}\bigg)^2\bigg(\frac{\xi\sin\theta}{\xi_0\sin\theta_0} \bigg)^2. 
$$
By solving this equation we get $\rho_i/\rho_e$ and hence the magnetic buoyancy ($M$).
In our simulation, we considered a thick accretion disc
around a black hole of mass $10~M_{\sun}$. To compute in a reasonable time
frame, we assume a small disc with the outer boundary at $200~r_g$. The actual size 
of the disc is much larger than what we are assuming. However, since we are only 
interested in the generic behaviour of the flux tubes close to the centrifugal barrier
we will continue with these typical parameters throughout our simulations. Second, at this 
distance, all random flux tubes entering far away would be expected to have a 
toroidal geometry. We inject these flux tubes from the radial grid boundary, i.e., at the $200~r_g$ 
near the equatorial plane. We consider an inviscid disc so that the angular momentum 
remains constant throughout. The specific energy of the injected matter remains 
constant since we have not considered any radiative cooling process. For the detail 
description of the numerical setup for the disc simulation refer to Deb, Giri, Chakrabarti (2016). 
 For schemes, basic properties of the code, and different test results, please refer to Harten 
(1983), RCM97 and Ryu et al. (1995).
In a transonic flow there are only two free parameters: angular momentum and energy. This is 
much less than any other solution, because the sonic point condition eliminates the need of supplying
any more free parameters. With these two parameters, we know both the sonic points of the flow, the
shock location and all other properties of the flow. In a Keplerian disk, the entire $\lambda(r)=
\lambda_{Kep} (r)$ is assumed without any reason, which is unacceptable. 
Every simulation solves a number of differential equations which must have the boundary condition. 
Otherwise, the solution cannot be found. We supply only $\lambda$ and energy as the boundary condition. 
In return we get the rich science of self-consistent advective flow parameters throughout the grid.
Angular momentum is defined as, $\lambda = rv_{\phi}$  in 
the unit of $2GM/c$. The values chosen are less than the value of $\lambda$ at marginally stable 
orbit, as per the theory of advective flow. A little higher than
the marginally stable value is also allowed (till marginally bound value of 2.0) 
in order to have a sonic point.  In Giri and  Chakrabarti (2013) it has been shown that in presence 
of viscosity, starting from angular momentum lower than  $\lambda_{ms}$ ,  Keplerian disk is formed. 
So, for inviscid case we chose $\lambda$ < $\lambda_{ms}$ so that the flow always remain sub Keplerian.
Total energy($\varepsilon$) of the injected flow is chosen to be less than the 
rest mass energy of the electron.
We use the injected flow to have constant specific angular momentum of (i) $\lambda = 1.6$, and 
(ii) $\lambda = 1.7$ and for each of these cases we use the specific total energy 
$\varepsilon = 0.001,~0.002,~0.006$.In future, we wish to include viscosity and create Keplerian disk as well on the equatorial plane.
 Magnetic flux rings at the outer boundary are placed near the equatorial plane ($\theta 
= 89^{\circ}$) with a initial magnetic buoyancy ($M_0$) which we calculate by taking the ratio 
between the magnetic pressure and external gas pressure. We have injected the flux tube only 
after the flow reaching an equilibrium configuration in order to remove the effects of transient 
phase of the simulation. We couple the equations of motions for the flux tube with the hydrodynamic 
TVD code. We modify the source function as given in eq. 4 by adding Lorentz force term. This will 
include flux tube's effect on the fluid. The input parameters, namely,  angular momentum and total energy 
gives unique injection velocity and sound speed at the outer boundary. These together with any 
density at the outer boundary which is scaled as unity at the equatorial region, gives the accretion rate. In a non-
dissipative flow, the result does not depend on the density explicitly. However, we calculate the 
density, velocity, pressure and temperature distribution  using time dependant TVD code based on the 
boundary values and the equation of state only. These flow parameters are used as the input parameters
for computing a flux tube's evolution inside the disk since the drag, buoyancy etc. depend on the 
environment in which the tube is moving and those in turn are plugged in the eqs. (7-9) and 
we numerically solve them using the fourth order Runge-Kutta method.   

\section {Results}

\subsection{Behaviour of the flux tubes inside an advective flow}

The trajectories of the flux tubes as obtained in our simulations 
are plotted in $r-z$ plane. Figure-1 shows the trajectories for the 
flux tubes injected with initial cross sectional radii $0.001,~ 0.005,~ 0.01, \& ~0.1~ r_g$ 
respectively released from the outer boundary with two different flow 
energies, namely, $\varepsilon = 0.001, 0.002$ (marked). The reasons behind choosing such 
small cross sectional radii are explained below. First, the axisymmetric flux tubes have to be made {\it ab initio} through shear and 
reconnection processes. So they are, by definition filamentary.  Second, flux tubes' cross 
sectional radii are chosen to be small compared to gravitational radius as well as scale 
height so that we can make the approximation that the variation of disk variables within 
a flux tube is negligible. Third, even if the filaments join and make tubes of thick cross
-sectional area, the drag force will be too high  and buoyancy force will remove them from the 
disk.  Angular momentum of the flow was chosen to be $1.6$.
 Near the axis since centrifugal force is very strong, a vortex like  opening  forms which is 
called the "chimney" or the funnel wall. We have observed that before getting accreted by the 
black hole flux tube undergoes oscillation very close to the black hole for a significant time
 and actually emerge in either of the funnels. For both the energies, flux tubes having initial 
$\sigma < 0.1 r_g$ emerge in the chimney (In our Figures, they appear as though they are entering into the 
black hole. But, in reality, they do not, and manage to escape). Flux tubes with $\sigma\geq 0.1 r_g$ are expelled 
for the lower energy case. As we increase the energy of the flow the flux rings tend to oscillate 
more as they will have more kinetic energy than lower flow energy configuration. Since this is 
an inviscid flow, Coriolis force will not play any part in this and flux tubes will move inwards along the direction of local pressure gradient.  

\begin{figure}
	\includegraphics[width=\columnwidth,trim= 0 .5 0 1.4, clip=true]{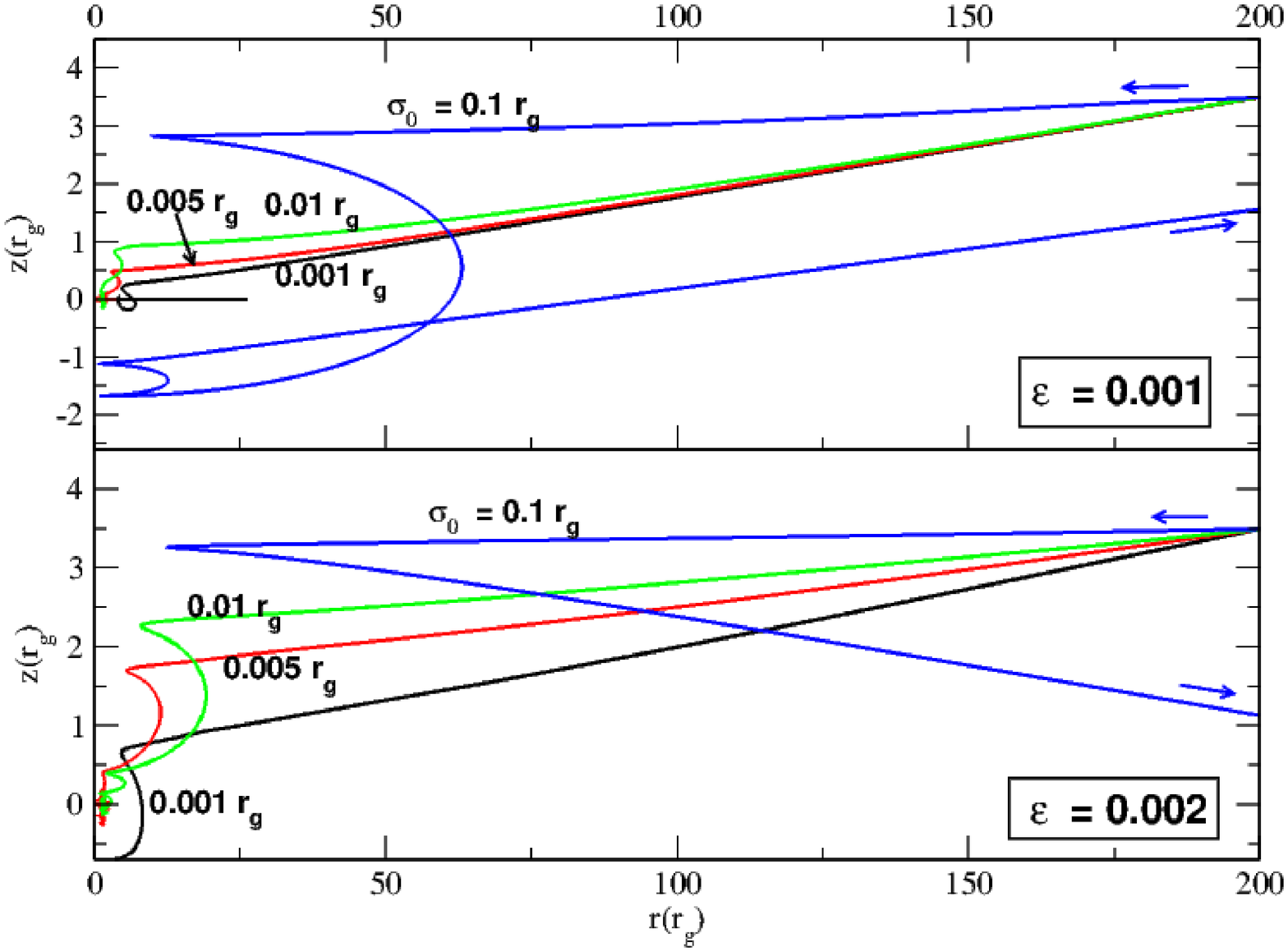}
    \caption{Trajectories of flux tubes injected from the outer 
         boundary i.e., $r=200 r_g$ and $\theta=89^{\circ}$ with zero initial 
         velocity. Trajectories are in $r = R\sin\theta$ vs. $z=R\cos\theta$ plane. 
         The trajectories are drawn for a flow with angular momentum $\lambda=
         1.6$ and energies $0.001$ (upper panel) and $0.002$ (lower panel).
         $\sigma$ is the cross sectional radii of the injected flux tubes.
         Here $\sigma$ values are $0.001~r_g$ , $0.005~r_g$, $0.01~r_g$ and $0.1~r_g$.}
    \label{fig:example_figure}
\end{figure}
\begin{figure}
	\includegraphics[width=\columnwidth, trim= 0 .5 0 1.4, clip=true]{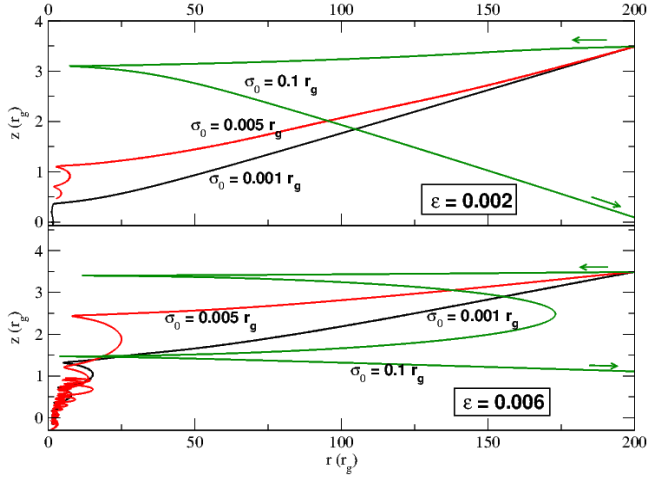}
    \caption{Trajectories of the flux tubes injected from the outer 
         boundary i.e., $r=200 r_g$ and $\theta=89^{\circ}$ with zero initial
         velocity. Trajectories are drawn in $r = R\sin\theta$ vs. $z=R\cos\theta$ 
         plane. The trajectories are drawn for a flow with angular momentum 
         $\lambda=1.7$ and energies $0.006$ (lower panel) and $0.002$ 
        (upper panel). $\sigma$ signifies the cross sectional radii of the flux 
         tubes for which the trajectories are drawn. Here $\sigma$ values are 
         $0.001~r_g$, $0.005~r_g$, and $0.1~r_g$.}
\end{figure}

\begin{figure}
	\includegraphics[width=\columnwidth, trim= 0 .5 0 1.4, clip=true]{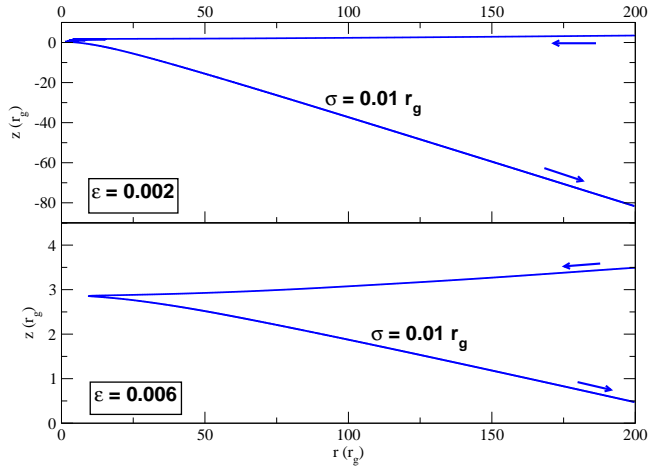}
    \caption{Trajectory of the flux ring having cross sectional radius
            ($\sigma$) $0.01~r_g$ released at $r=200 r_g$ and $\theta 
            = 89^{\circ}$ with zero initial velocity. Angular momentum of
            the flow is $1.7$ and the flow energies are $\varepsilon=0.002$ 
            (upper panel) and $0.006$ (lower panel).}
\end{figure}

We repeat the simulation for the same set of $\sigma$s and for angular momentum 
($\lambda = 1.7$) and the energies $\varepsilon = 0.002,~0.006$ for which the 
flux tube's dynamics are drawn in fig. 2 and fig. 3. From Fig. 2 
and Fig. 3 it is quite evident that flux tubes having $\sigma$ less than
of $0.005~ r_g$ will fall onto the black hole which is same as what we have 
seen in Fig. 1 but contrary to the previous case (Fig. 1) in this case, the
flux tubes with $\sigma \geq 0.01~r_g$ will be expelled. The flux tubes with 
different injected $\sigma$ will take different time to complete the dynamics.
In case of Fig. 1, for $\varepsilon=0.001$, end times for flux tubes with initial
cross sectional radii $0.001~r_g$, $0.005~r_g$, $0.01~r_g$ and $0.1~r_g $ are $t_{end}=~2.12,~
2.7,~9.46,~\&~5.8$ seconds respectively and for $\varepsilon=0.002$, $t_{end}=~2.2,
~19.8,~20.4,~6.1$ seconds in the same order. In case of Figs.-2 and 3 this $t_{end}$ 
is given as, for $\varepsilon=0.002$, $t_{end}=~8.65,~9.2,~3.34,~\&~5.77$ s, 
and for $\varepsilon=0.006$, $t_{end}=~21.2,~22.3,~5.15,~\&~12.64$ s. It is 
seen from our simulation results that the flux tubes with oscillatory feature has 
longer residence time inside the disc. \textbf{The unit of time used for the purpose of simulation
is $2GM/c^3$. After the simulation is completed in dimensionless units, we use this factor to 
convert time in physical unit (seconds). Mass of the black hole is chosen to be
$M = 10 M_{\odot}$).} The infall time in the simulations is $\sim 0.6 -0.7$ s
and it can be observed that the residence time is several times longer than this for all the cases
we investigated.

\begin{figure}
\includegraphics[width=\columnwidth, trim= 0 .5 0 1.4, clip=true]{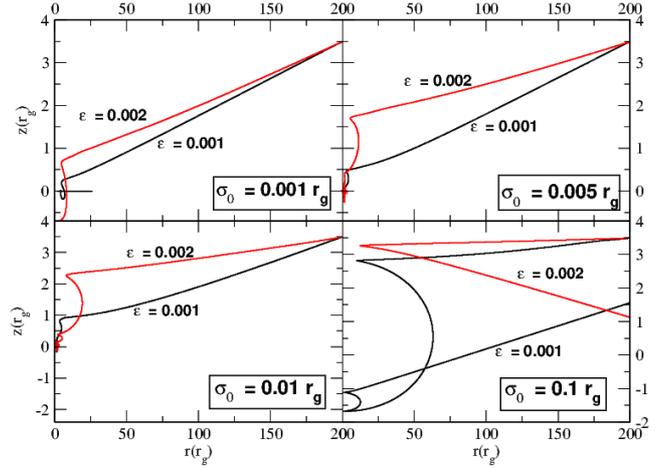}
\caption{Trajectories of flux tubes having same $\sigma$ but with different 
         flow energies (marked) are drawn to demonstrate energy dependence 
         of the flux tube's path inside the disc. $\varepsilon = 0.001 \& 0.002$ are
marked on the curves. Angular momentum of the flow is $1.6$.}
\end{figure}
Figure 4 and Fig. 5 demonstrate the energy dependence of the trajectories of the flux tubes
released in a flow having angular momenta $1.6$ and $1.7$ respectively. As we increase the 
energy, the flow becomes more turbulent and thus it imparts more kinetic energy to the
flux tubes. In Fig. 4 we have considered the flow energy to be $\varepsilon = 0.001~ \&~0.002$ 
and in Fig. 5 energy is $\varepsilon=0.002~\&~0.006$.
\begin{figure}
\includegraphics[width=\columnwidth, trim= 0 .5 0 1.4, clip=true]{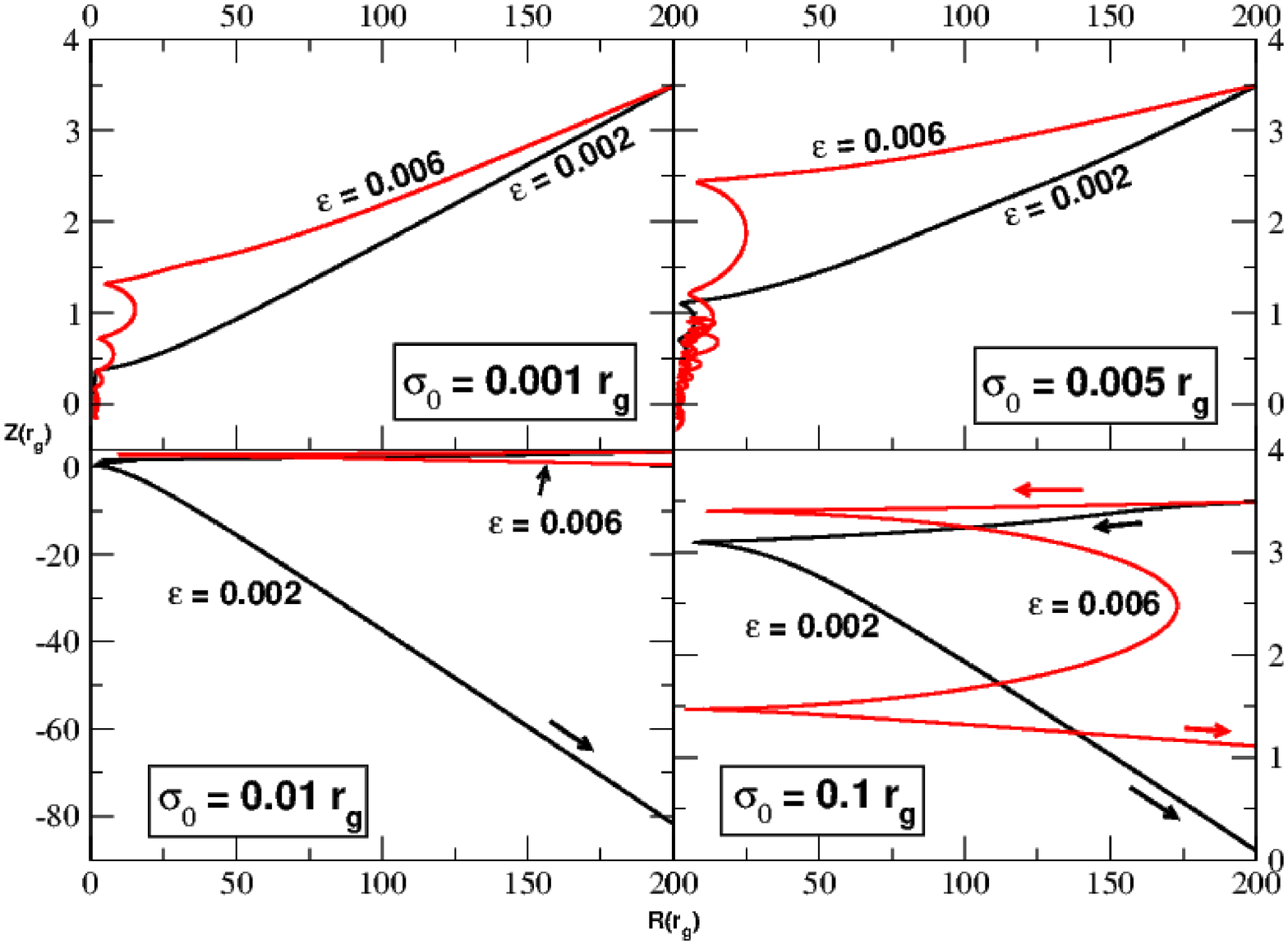}
\caption{Same as in Fig. 4 except $\varepsilon = 0.002 \& 0.006$ and specific angular 
              momentum of the flow is $1.7$.}
\end{figure}

As the flux tube moves in, the direction of the pressure gradient force it tries to maintain
pressure equilibrium (i.e., an equilibrium between the external gas pressure and internal gas
pressure together with the magnetic pressure) with its immediate surrounding. When the flux 
tube contracts during its journey towards the black hole, it will cause the internal density to 
increase and tries to become heavier than the surrounding. However, near the black hole, 
density of the gas in the disc also rises rapidly in the same directions. The result is that
the inward journey of the flux tube is halted and the tube bounces back. 
During ensuing outward motion, the flux tube becomes relatively less 
dense very rapidly and depending on the direction, it can totally escape or 
it can start to oscillate depending on the relative change in the  flux cross-section and the
disc density both of which control the buoyancy force. As in the Sun, where the flux tubes 
are anchored in the region between the radiative and convective zones and at the base of the 
anchored flux tubes entropy gradient changes sign, in case of the disc, similar
anchoring effect of the oscillating flux tubes could be expected under some circumstances.
In the Sun, the anchored flux tube emerge on the surface due to Parker instabilities
and similarly the the anchored flux tubes may also appear close to the chimney due to some 
perturbative effects causing magnetic activities (collimation 
and acceleration of jets) (see, CD94(I) and CD94(II)). However, demonstration of the 
instabilities requires a three dimensional simulations which is beyond the scope of this paper.

In a turbulence free laminar perfect flow, the entropy is expected to be constant. 
However, in our case, due to the tag-of-war between the gravitational force and 
the centrifugal force the flow becomes turbulent near the centrifugal barrier and could 
even form shocks. This will generate entropy. In Fig. 6, we plot the map of the radial 
component of the entropy gradient ($\vec{\nabla}_{r}$) 
at two different times, $t\sim~4.06$s and $6.33$s respectively. The angular momentum ($\lambda$) and 
specific energy of the flow are considered to be $1.6$ and $0.002$ respectively. In both the panels
we see that the $r$ component of the entropy gradient changes sign i.e., it goes from postive 
to negative and vice versa which generates a Solar interior like situation where the oscillating flux 
tube can be provided an anchorage by the entropy gradient.

\begin{figure}
\includegraphics[width=\columnwidth, trim= 0 .5 0 2, clip=true]{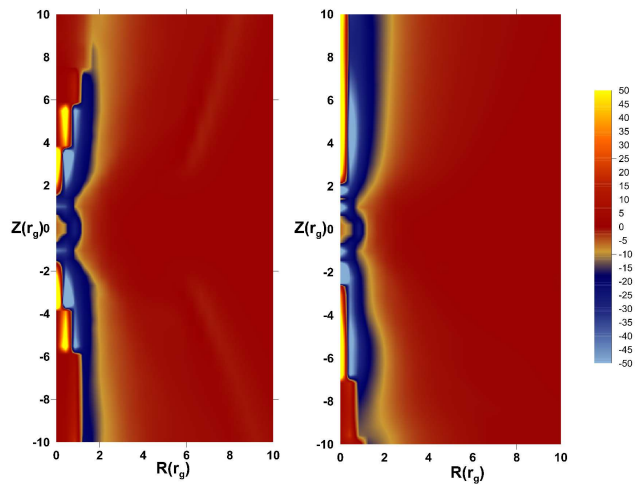}
\caption{Time variations of radial component of the entropy gradient plot of the flow having 
         angular momentum ($\lambda$) $1.6$ and energy ($\varepsilon$) $0.002$.
         Both the plot shows that the radial component of the entropy gradient
         switches sign from positive to negative and vice versa. This switching
         is responsible for providing an anchorage of the oscillating flux
         tube and consequently may cause a corona like structure. Two plots are drawn 
         at $t\sim~4.06~\&~6.33$ s respectively.}
\end{figure}
Figures 7 and 8 show variations of the cross sectional radius of the magnetic flux tubes as a function of
the vertical distance the flux ring traverses. The panels in each Figure contain the $\sigma$-
variations for different initial cross sectional radii ($\sigma_{0}$). The initial $\sigma$s for the
plots are $\sigma_{0}~=~0.001,~0.005,~0.01,~\&~0.1 r_g$. All the Figures are drawn for different angular 
momentum and specific energy of the flow.
Figures 7 and 8 have the $\lambda$ and $\varepsilon$ 
configuration as $(\lambda,\varepsilon)= ~(1.6,~0.002)~\&~(1.7,~0.006)$ respectively. The divergence of 
the magnetic field must always
be zero and the net flux of the toroidal flux tubes remains constant throughout its flight. 
Thus, when $\sigma$ decreases, the magnetic field intensity ($B$) will 
increase $(B\propto \frac{1}{\sigma^2}$ and vice versa, affecting the buoyancy force.  

\begin{figure}
\includegraphics[width=\columnwidth, trim= 0 .5 0 1.4, clip=true]{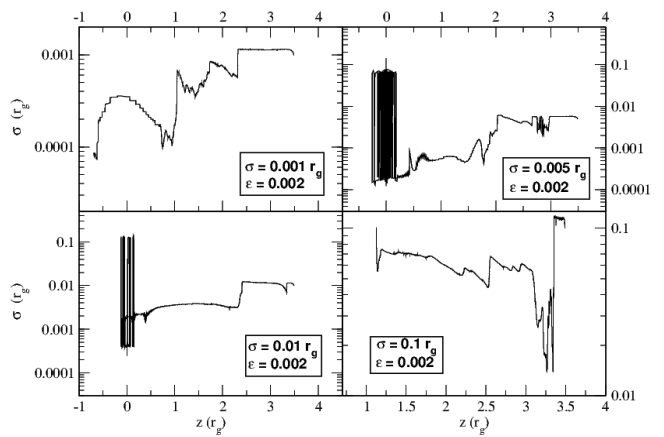}
\caption{$z$ variation of the radius of cross section($\sigma$) of flux tubes
         released in an inviscid accretion flow with energy $\varepsilon=0.002
         $ and angular momentum($\lambda$) $1.6$. Each panel 
         shows the $\sigma$ variation for different initial cross sectional
         radius. initial $\sigma$'s are $\sigma= 0.001,~0.005,~0.01,~\&~0.1
         ~r_g$}
\end{figure}
In Figs. 7 and 8 it can be observed that the cross sectional radius rapidly increase and decrease 
thus to conserve the flux, the field will change inverse squarely with the cross-section. Typically 
when the flux tube goes inward the field becomes intense on an average (barring oscillations) and 
thus the magnetic pressure holds matter inside and does not allow them to leak out side ways. Now, 
There are two opposing effects: While moving in, the flux increases and collimates outflowing matter 
at the base of the jet. However, as the flux tube leaves along the axis its pressure falls and 
its ability to collimate decreases, although not as much, since the jet itself is becoming lesser 
dense as it expands out. We see the average effect rather than a sustained effect. If thousands of 
such tubes could be injected, they would have a sustained effect in collimating the jets.
\begin{figure}
\includegraphics[width=\columnwidth, trim= 0 .5 0 1.4, clip=true ]{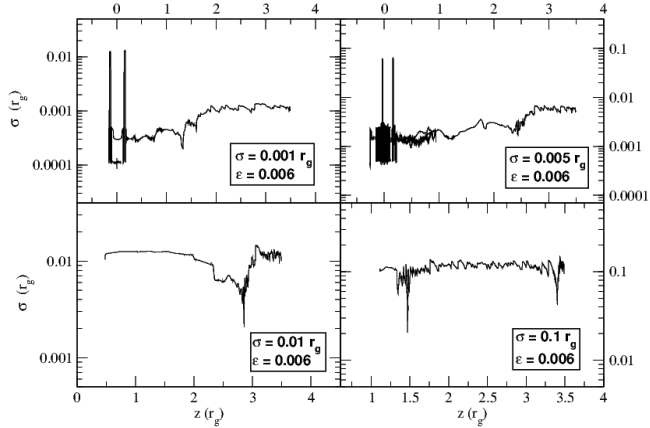}
\caption{$z$ variation of the radius of cross section ($\sigma$) of flux tubes
         released in an inviscid accretion flow with energy $\varepsilon=0.006$ 
         and angular momentum ($\lambda$) $1.7$. Each panel 
         shows the $\sigma$ variation for different initial cross sectional
         radius. initial $\sigma$'s are $\sigma= 0.001,~0.005,~0.01,~\&~0.1
         ~r_g$.}
\end{figure}

\subsection{Effects on jets and outflows}

There are two types of jets: (a) sustained slow moving outflow which is always coming out of the 
post-shock region and which are collimated by the flux tubes on an average and (b) blobs of fast 
moving matter which are squirt out due to sudden collapse of the inner region of the disk. These are 
due to  magnetic tension. If the flux tube were very strong, it would collapse due to tension 
destroying the CENBOL and produce blobs by the so-called magnetic rubber band effects (b above). 
These blobs are fast moving to begin with. The type (a) outflows discussed above are accelerated indirectly: the collimation of the jet reduces its lateral expansion retaining its initial energy. The 
cross-sectional area increases slowly with distance (along Z axis) and thus they are accelerated.
\begin{figure*}
\includegraphics[width=\textwidth, height=\textwidth]{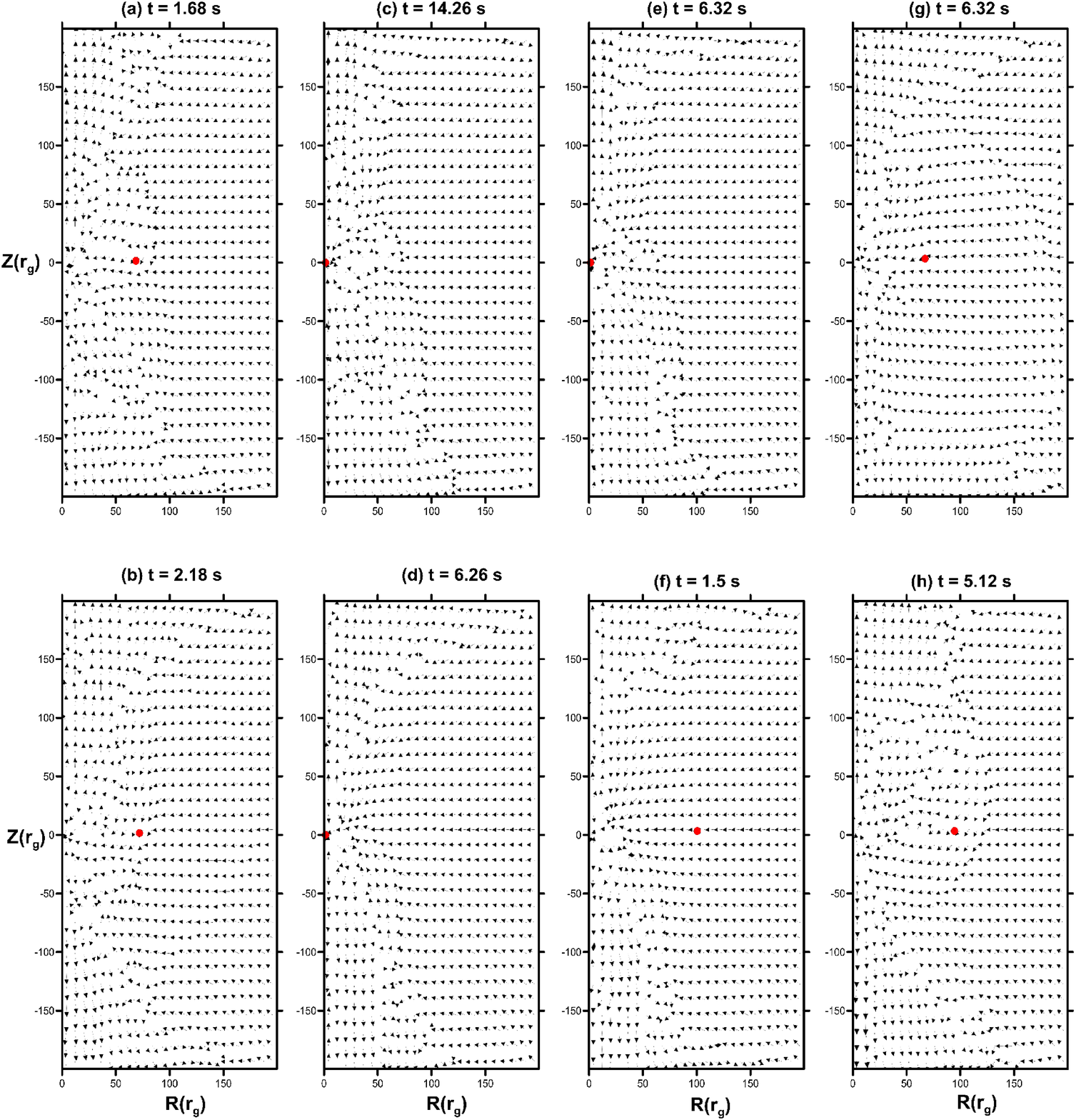}
\caption{Velocity vector plot of the flow with magnetic flux tubes. (a,b), (c,d)
, (e,f), and (g,h) are velocity vector fields of the flow having magnetic flux 
tube of cross sectional radii $0.001,~0.005,~0.01,~0.1~r_g$ respectively. Angular momentum and specific energy are $1.6$ and $0.002$ respectively. The times specified  are the same as in Fig. 10. The dots signify the position of 
flux tube at the respective times specified in each panel.}
\end{figure*}

\begin{figure*}
\includegraphics[width=\textwidth,  trim= 0 .5 0 2, clip=true]{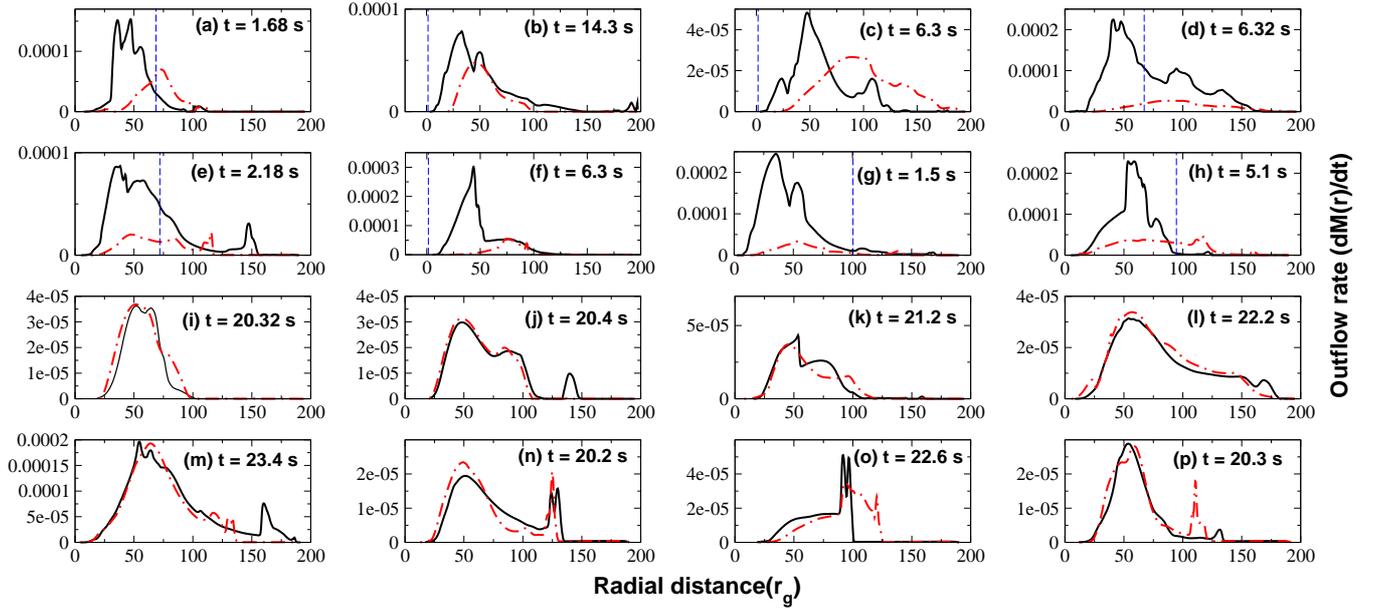}
\caption{Radial distribution of the outflow rate ($\dot{M}_{out}$) of the 
         flow having the specific angular momentum ($\lambda$) $=1.6$ and energy 
         ($\varepsilon$) $=0.002$. The black solid curve represents the 
         outflow rate for the flow with magnetic field and red solid curve 
         (dot-dashed in hard copies) denotes the result in non-magnetic case. 
         The upper two rows (a-h) of the plot show the collimation of the 
         outflow from upper and lower quadrants respectively for different 
         flux tubes with different $\sigma$.  The lower two rows (i-p) depict 
         the dissipation of the collimating effect once the flux tube has 
         escaped or fallen into the black hole. The vertical dashed lines drawn
         in panels of first two rows depict the position of the flux tube at 
         time for which the outflow rates are drawn.}
\end{figure*}
\
In Figs. 10 and 11 we plot radial distribution of the outflow rate obtained from both the 
quadrants, upper and lower. Here, we have compared the outflow obtained from the flow with the 
magnetic flux tube injected in it with the flow that does not have any presence of magnetic flux tube
in it. The first two rows bear the signatures of the outflow being collimated in the presence of the
magnetic flux tube while the lower two rows signify fading away of the collimating effects of 
the flux ring as it escapes from the system or falls onto the black hole. The black curve depicts the 
outflow rate for the magnetic case and the red curve is for the non-magnetic case. Both the Figures
(Fig. 10 and Fig. 11) are drawn for a similar set of $\sigma~(=~0.001,~0.005,~0.01,~0.1~r_g)$.
We have run our simulation for $t\sim
23.76~s$ and in case of both the Figures each panel represents the snapshots of outflow rate variation
at different times (see, Figs. 10 and 11). There was no specific reasons for choosing these particular times. We investigated and compared outflow profiles continuously. We want to show the effects 
of flux tubes when it is present in the flow and how its effect diminishes when it leaves the flow. So 
we covered the entire run time and presented four pictures at each time where aforementioned effects 
were prominent enough. 
\begin{figure*}
\includegraphics[width=\textwidth, trim= 0 .5 0 2, clip=true]{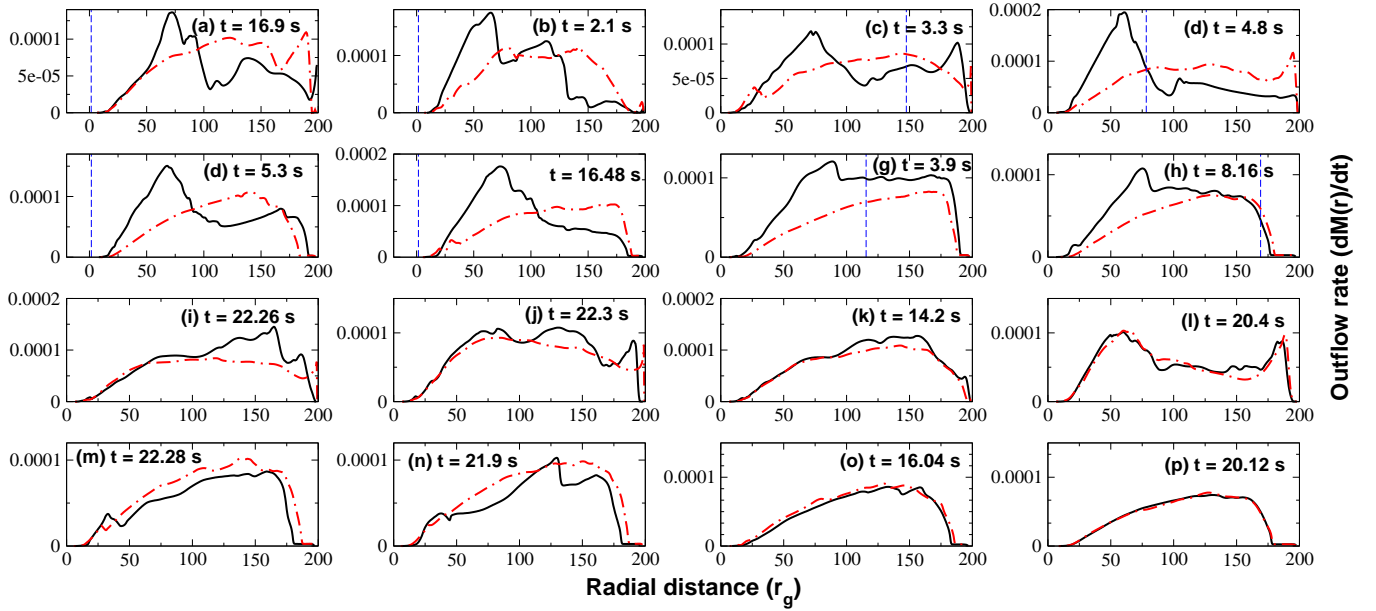}
\caption{Same as Fig. 10, but $\lambda = 1.7$ and $\varepsilon=0.006$.}
\end{figure*}

The angular momentum ($\lambda$) and the specific energy ($\varepsilon$) for Fig. 10 are chosen to be
$1.6$ and $0.002$ respectively and for Fig. 11 the value of $\lambda$ and $\varepsilon$ are $1.7$ and
$0.006$ respectively. In Fig. 10 (a-h), the collimation of the outflow is prominent. The
outflow rate for the magnetically confined flow has a sharp peak at around the region of 
$30-50~r_g$. However, the outflow rate for the non-magnetic flow achieves a maximum value at 
around $50-90~r_g$. Also the maximum value of the outflow rate is much higher, almost $2-3$ fold higher
than the maximum value of outflow rate for the non-magnetic case. 
In contrast to what we have seen in Fig. 10, in case of Fig. 11, 
in the panels (a-h) we observe that the collimating effect is not as
prominent as it is in the previous case. In this case, the outflow rate for magnetic flow reaches the 
maximum at around $50-80~r_g$ which is farther than what we have seen in Fig. 10(a-h). The reason clearly 
lies in the fact that the higher angular momentum caused high value of centrifugal force in Fig. 11, and thus 
it is difficult to collimate the flow by the field lines of similar initial strength.
In Fig. 10(i-p), the outflow rate is plotted after the magnetic flux tubes have either
escaped from the simulation box or fallen onto the black hole. We see that the effects of the magnetic flux tubes
have started to fade away. In case of Fig. 10 (i, j, m, and n), the outflow rate is plotted long after the
flux tube has escaped or fallen onto the black hole. In these cases, it is evident that
the effects of the flux tubes have faded away significantly and the
outflow rate of magnetic and non-magnetic case almost match with each other. 
On the other hand, since for the case depicted in Fig. 10 (k) and (o),
the radial variation of the outflow rate is plotted 
only a few tens of dynamical time 
after the flux tubes escape from the simulation box, the outflow rates of both 
magnetic and non-magnetic cases do not match with each other but the trend of doing so is evident.
In Fig. 11 (i-p), the fading away of the effects of magnetic field is very much evident 
for the flux tubes with high injected $\sigma$ and in the cases 
Fig. 11(k, o, l, p), the outflow rate for both magnetic and non-magnetic 
cases nearly match with each other.
\begin{figure*}
\includegraphics[width=\textwidth, trim= 0 .5 0 2, clip=true]{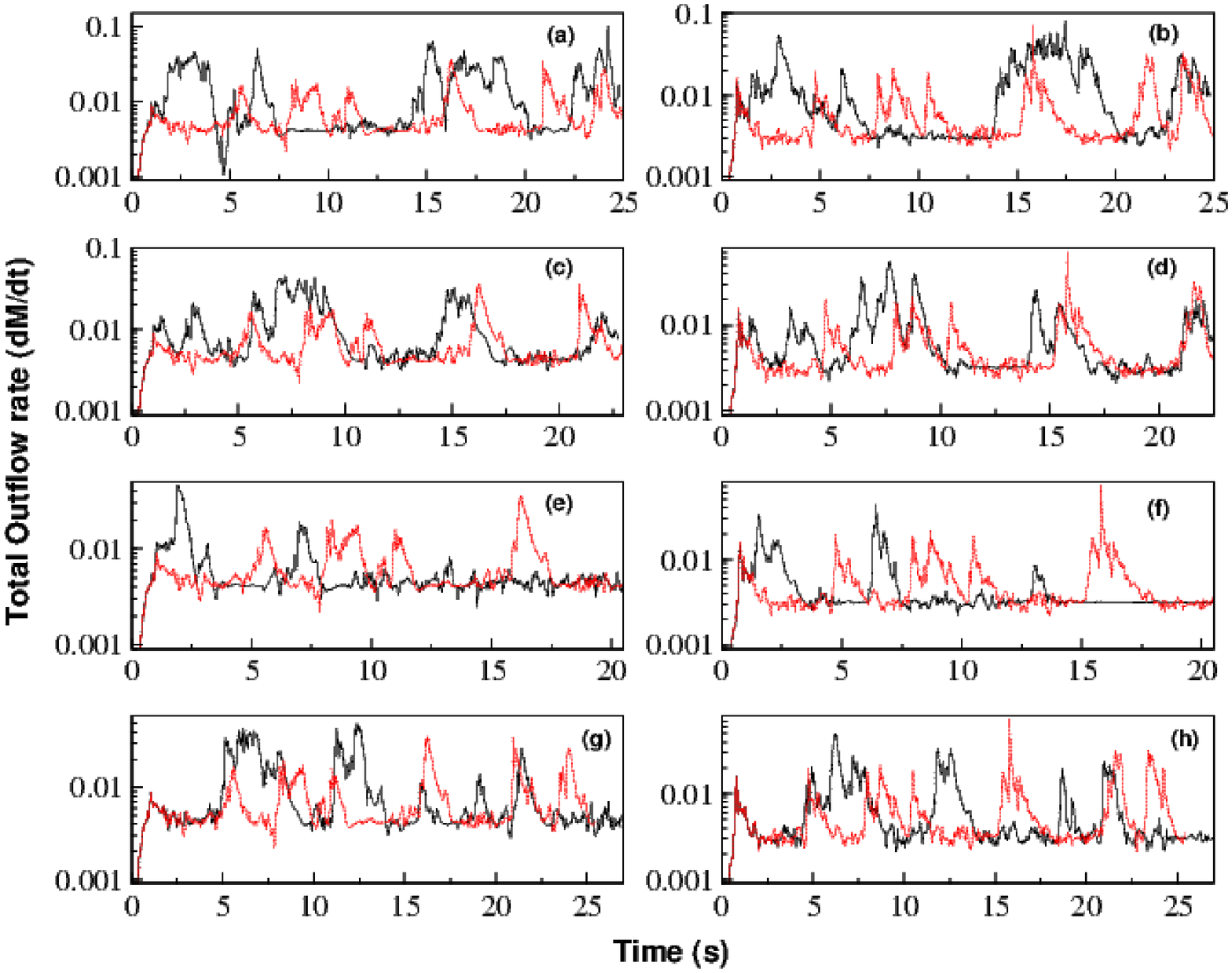}
\caption{Time variation of total outflow rate for magnetic and non-magnetic 
cases with angular momentum $1.6$ and specific energy $0.002$. Panels (a,c,e,g)
represents net outflow rate for upper quadrant and panels (b,d,f,h) represents
total outflow rate for lower quadrant of the flow. Solid line represents total 
outflow rate for magnetic cases and dashed line represents the non magnetic 
cases. Panels (a,b), (c,d),(e,f), and (g,h) are drawn for flux tube with cross 
sectional radii $0.001,~0.005,~0.01,~0.1~r_g $.}
\end{figure*}
 Figure-12 shows time variation of the total outflow rate for upper and 
lower quadrants. Outflow rate increases when the magnetic flux tube remains in 
the flow but when it leaves the system, the outflow rate returns back to that of 
of the non-magnetic flow. 

As the outflowing matter gets pinched and squirts off along the vertical direction due to presence of 
the magnetic flux tube, the z-component of the velocity of matter is expected to increase significantly in comparison to the non-magnetic scenario. This signifies that in the presence of the toroidal
flux tube, the outflowing matter gets accelerated. To show that we captured this effect also, in Figs. 13 and 14 we plot
contour maps of the  differential velocity, i.e., the difference of z-component of velocity between magnetic and non-magnetic case. 
\begin{figure*}
\includegraphics[width=\textwidth, trim= 0 .5 0 2, clip=true]{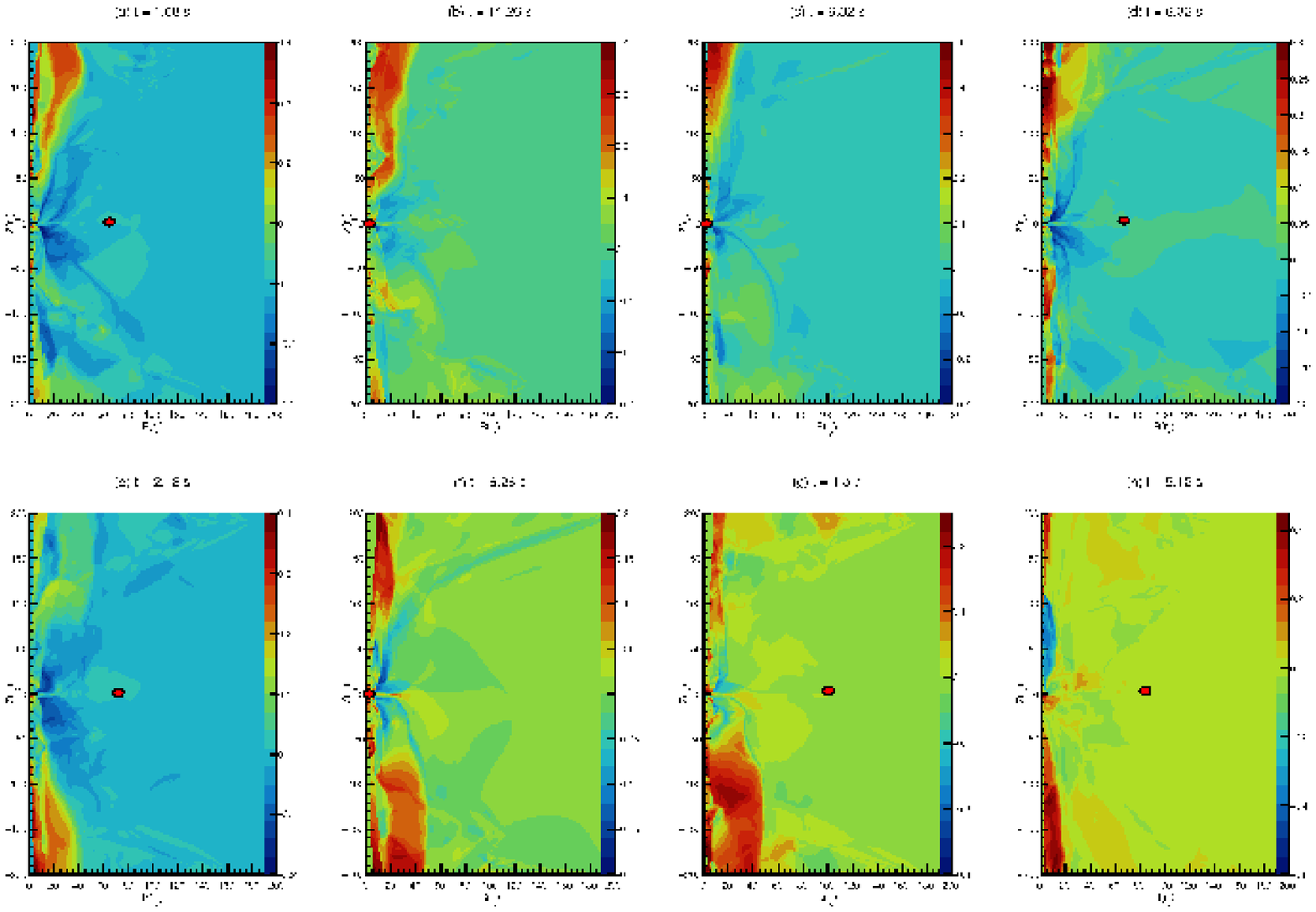}
\caption{Map of the difference between z-velocity of magnetized and 
         non magnetized flow. (a-d) represent the upper quadrant and (e-h) represent 
         the lower quadrant of a two quadrant flow. Each pair of panels (upper 
         and lower) represent different cross sectional radius. Here,$\sigma~
         (=~0.001,~0.005,~0.01,~0.1~r_g)$. Angular momentum($\lambda$) is $1.6
         $ and specific energy ($\varepsilon$) is $0.002$. Each panel is drawn 
         for different times were as the Fig. 10. The circles drawn in the panel
         give the position of flux tube at times for which the panels are drawn.          }
\end{figure*}

In Figs. 13 and 14 upper panels show that the acceleration is prominent 
in the upper quadrant and the lower panels show that the acceleration is prominent in lower quadrant of a two quadrant flow. Each of the Figs.13(a-h) and Figs.14(a-h) is
drawn for different cross sectional radii of flux tubes. Here, $\sigma$s are the usual set for which the other Figures are drawn. 
The times for which the maps are drawn, are the same as mentioned in the Figures 
showing collimations of the outflow. We plotted $(v_{z,mag}-v_{z,non
mag})$. If this difference is positive that will mean that due to the 
presence of magnetic flux tubes the z-component of the velocity is increased 
i.e., the flow it has been accelerated.
Figure 13 is drawn for $\lambda=1.6$ and Fig. 14 is drawn for $\lambda=1.7$.
In Fig. 13, it can be observed that for all the cases the z velocity of magnetized flow has increased significantly within the radial distance of $5-50 r_g$ for both upper and lower quadrant and the maximum value of the velocity difference has gone up to $0.36c$ in some cases. 
In contrast to Fig. 13, in Fig. 14, the region where the z-velocity 
of magnetized flow increases is broadened and it stretches up to 
the radial distance $100~r_g$ and also in this case, 
the maximum value of the velocity difference that can be achieved 
is $0.18c$ which is much lower than what is seen in Fig. 13. 
This disparity is due to the fact that for higher angular 
momentum the centrifugal force becomes higher which causes 
less collimation as seen in the Fig. 10 and thus acceleration of jets/outflows 
by the field lines injected with similar initial strength was reduced. 
  
\begin{figure*}
\includegraphics[width=\textwidth, trim= 0 .5 0 2, clip=true]{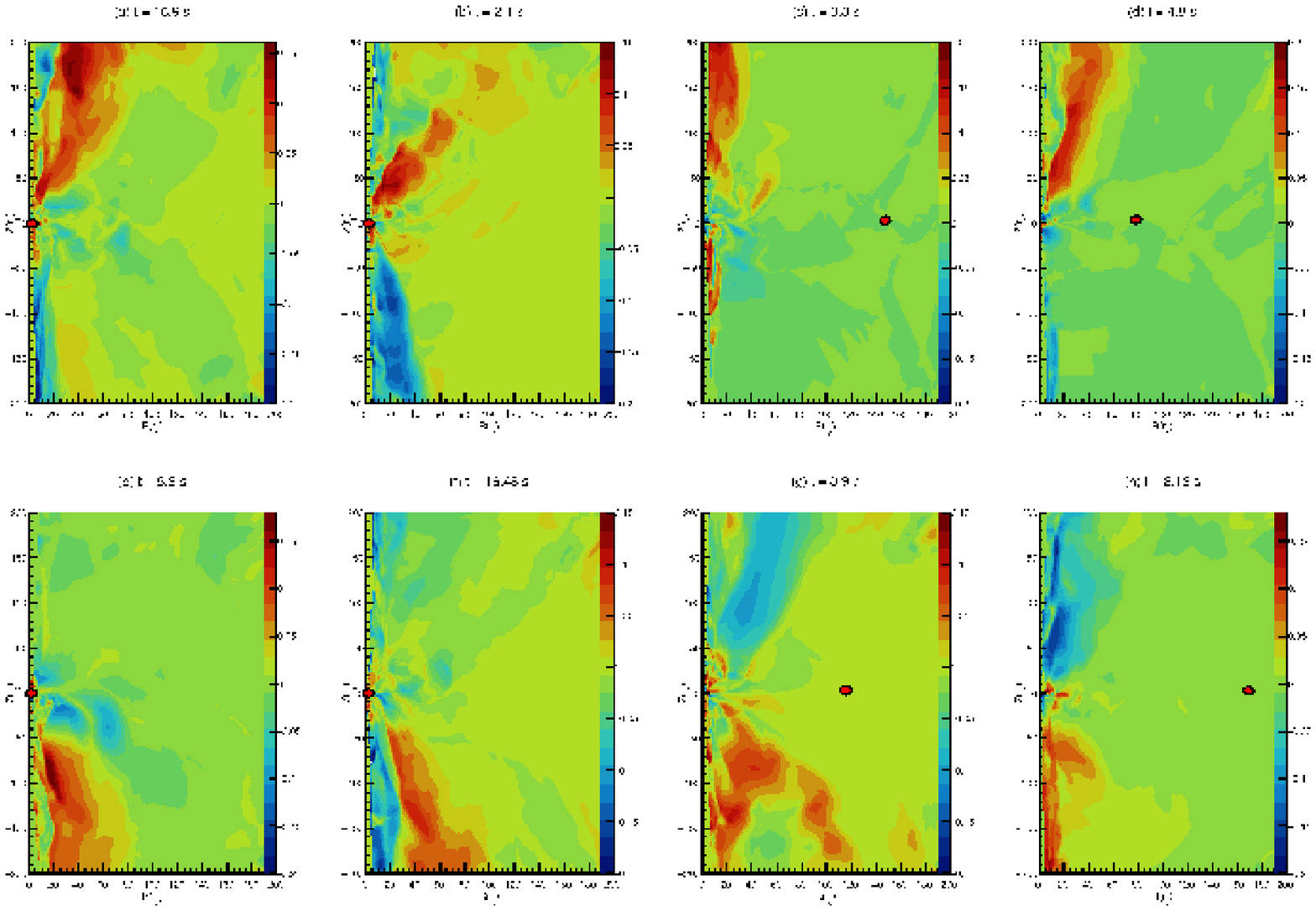}
\caption{Same as Fig. 13, but $\lambda = 1.7$ and $\varepsilon=0.006$. The time 
         written in each panel is the same as Fig.11}
\end{figure*}

\section{Discussions}

In this paper, we studied the dynamics of magnetic flux tubes which 
are released at the outer edge of a time dependant two quadrant 
thick advective disc and their role in collimation and acceleration of the jet and 
outflow from the upper boundary. In earlier studies, 
such as CD94(I), CD94(II), simulations were carried 
out to study the dynamics of flux tubes in the realm of time independent 
thick disc (Paczy\'nski \& Wiita 1980, Rees et al. 1982, Chakrabarti 1985). 
In those earlier works, the general conclusions drawn were that the flux tubes released between the
inner edge and the centre of the disc emerge in the chimney irrespective of the angular momentum 
distribution, accretion flow rate, cross sectional radius, and magnetic intensity of 
the flux tubes making 'chimney' of the funnel wall, magnetically the most active region. Flux rings released outside 
the centre of the disc may or may not emmerge into the chimney depending upon the interplay among 
the several disc and flux tube parameters.  We observed the emergence of the flux tubes in the funnel wall or upper boundary. However,
it was also possible to construct physical models of time independent thick accretion discs 
that can provide the storage of weaker flux tubes that instead of being expelled away tend to oscillate around equipotential surfaces until they get amplified and buoyant and leave the system. In an effort to study 
more realistic cases, in the present work, we followed the dynamics of the flux tubes released in a 
time dependent accretion after removing the reflection symmetry (Deb et al. 2016) condition 
as used in earlier purely hydrodynamic simulations (GC10). 
We also answered whether these magnetic flux tubes aid in the acceleration and collimation of the 
jets or not. In order to do this we have injected a single flux tube in each simulation, at the outer 
boundary of the disc after a few dynamical time so that the initial transient phase of the flow is 
be over and the flow will settle down to a stationary solution. As the rotational velocity 
becomes stronger when the flow approaches the black hole, only dominant component is expected to be the 
azimuthal component. Hence, without any loss of generality we inject toroidal fields. It is observed 
from our simulation that depending upon the initial cross sectional radius of the flux tube and the 
flow parameters such as angular momentum and energy, flux tubes can move directly towards the chimney 
or oscillate till it is expelled away. In a low angular momentum accretion flow from winds of 
companion, the angular momentum could be from $\sim 0$ to $\sim 2$ without much instabilities. However, the shock location becomes higher for higher angular momentum and thus the amplification of the flux 
tubes are larger and have chances for ejection of flux tubes earlier on. The same point goes of 
specific energy (in units of $c^2$). For a hot  advective flow, the specific energy (other than the 
rest mass=$1$), 
increases with increasing its temperature and speed. As Chakrabarti (1989) showed, the shock
location increases with energy. Thus raising both energy and specific angular momentum, the 
post-shock region forms farther out and makes the magnetic field difficult to be advected in
since the drag force considerably increases in regions of high density in the post-shock region.
Increase in magnetic field cross section would increase in buoyancy force as well.  
We also find that in case of certain angular momenta and energies, (i) the outflow rates ($\dot{M}
(r)$) from both upper and lower quadrants increase significantly in comparison to the outflow rates 
with respect to the non-magnetic cases (ii) The outflow rate is 
reaching its maximum value at much smaller radius, i.e., the spread of the outflow
at the upper and lower boundaries has reduced significantly. It is to be 
noted that though we do not see much fluctuations on a day to day basis in the observed jets, at the base, the fluctuations are natural since the inner edge could be oscillating and produce jets of sporadic rates. Soon after launching this fluctuations average out and we see only average effects far away. There are many instabilities in the disk, especially in presence of  a single field line, which is either in the upper half or in the lower half at a time, the configuration is disturbed and the fluctuations are expected. In presence of multiple field lines, distributed in both halves, such wild fluctuations should be averaged out. We note that the pinch felt by the outflow increases its outer velocity as well.

 In the beginning of the outbursts, the shock locations are higher and above discussion suggests that the jets would be weakly collimated at the the base. If the magnetic field in the companion is strong
so that the disk intersects it and amplifies and produces more flux tubes, we can expect stronger collimation of outflows in such system. If the companion star is non-magnetic and the field 
can only enter sporadically, the outflow will not be well collimated. Though we studied one tube 
at a time, our motivation is to understand what happens to the flow when an ensemble of flux tubes 
enter the disc, which is possibly the case in the realistic scenario. In that case, we will expect a 
faster and better collimated jet in a sustained manner. Thus we have a clear prediction that magnetic 
activity of the companion (or, the surroundings in case of super-massive black holes) is directly correlated with the emanation of stronger and collimated jets from the inner regions of the disc.

In case of the Sun, magnetic flux tubes are known to be anchored between the radiative core and 
convective envelope and they come out to the surface due to Parker instabilities. This is possible since 
the time scale of instability is much shorter as compared to the buoyancy time scale. However, in 
case of thick flows around black holes, pressure gradients are very strong and the flux tube may 
escape as a whole, especially those with stronger fields. By "escaping the
 disc" we mean that 
when the flux tubes enter into the funnel (Chimney) or leave the upper computational 
grid we assume that they escaped the disk. However due to topological constraints, they are 
not destroyed and will collimate the jet. If the jet is not formed, then they can move sideways and 
leave the system altogether. In the paper, we showed that as long as the flux tube is within the grid, 
the outflow is collimated, and its speed is higher. When the flux tube leaves the disk, the outflow 
relaxes back to the original shape.  Of course some flux tube may still 
pop-up and produce corona, but the probability does not seem to be strong, judging by our simulation
results. Furthermore, presence of small scale turbulence may tear off fields of larger $\sigma$ into 
smaller ones which then move in further. These flux tubes may also be responsible for a large number 
of astrophysical processes, such
as the variability of blazars, magnetic winds, production of high energy particles in coronae through Fermi acceleration 
processes etc. (CD94(II)). In many objects such as, GRS 1915 + 105 the 
variability classes namely $\chi_{1},~\chi_{3}$ and $ \beta $ are found to be associated with strong radio jets 
(Nandi et al. (2001), Naik \& Rao (2000), Vadawale et al. (2003), Vadawale et al. (2001) ). In 
case of $\beta$ class it is suspected that magnetic tension 
in the post-shock region 
becomes the most dominant component causing an abrupt collapse of this region.
This may signify that the magnetic field causes huge acceleration of jets 
(Nandi et al. (2001),
Naik \& Rao (2000),Vadawale et al. (2003), Vadawale et al. (2001) ). 
Detailed study of acceleration of jets is out of the scope 
of this paper but this will be discussed elsewhere.
We observed that only these initially filamentary flux tubes which are produced due to the 
presence of shear in the disc could be advected to the innermost regions of the disc. It is 
not impossible that many of such filaments merge due to higher density and make stronger flux 
tubes which then suddenly collapse and remove the inner region altogether. The opposite would 
be true when small scale turbulence is strong. These aspects will be dealt with in future.
%





\begin{thebibliography}{99}
\bibitem[]{} Batchelor G. K., 1967, An Introduction to Fluid Dynamics. Cambridge Univ.
Press, Cambridge
\bibitem[]{} Blackman E. G., 1996, ApJ, 456, L87
\bibitem[]{} Blandford R. D., Payne D. G., 1982, MNRAS, 199, 883
\bibitem[]{} Camenzind M., 1989, in Belvedere G., ed., Astrophysics and
Space Science Library Vol. 156, Accretion Disks and Magnetic
Fields in Astrophysics. pp 129–143, doi:10.1007/978-94-009-
2401-714
\bibitem[]{} Chakrabarti S. K., 1985, ApJ, 288, 1
\bibitem[]{} Chakrabarti S. K., 1986, ApJ, 303, 582
\bibitem[]{} Chakrabarti S. K., Bhaskaran P., 1992, MNRAS, 255, 255
\bibitem[]{} Chakrabarti S. K., D'Silva S., 1994, ApJ, 424, 138
\bibitem[]{} Chakrabarti S. K., Rosner R., Vainshtein S. I., 1994, Nature, 368,
434
\bibitem[]{} Choudhuri A. R., Gilman P. A., 1987, ApJ, 316, 788
\bibitem[]{} Coroniti F. V., 1981, ApJ, 244, 587
\bibitem[]{} D'Silva S., Chakrabarti S. K., 1994, ApJ, 424, 149
\bibitem[]{} Deb A., Giri K., Chakrabarti S. K., 2016, MNRAS, 462, 3502
\bibitem[]{} Eardley D. M., Lightman A. P., 1975, ApJ, 200, 187
\bibitem[]{} Fendt C., Camenzind M., 1996, A\&A, 313, 591
\bibitem[]{} Ferriz-Mas A., Schuessler M., Anton V., 1989, A\&A, 210, 425
\bibitem[]{} Fukue J., 1982, PASJ, 34, 483
\bibitem[]{} Galeev A. A., Rosner R., Vaiana G. S., 1979, ApJ, 229, 318
\bibitem[]{} Giri K., Chakrabarti S. K., Samanta M. M., Ryu D., 2010, MN-
RAS, 403, 516
\bibitem[]{} Giri K.,2015,Numerical Simulation of Viscous Shocked Accretion Flows Around Black Holes, by K. Giri. Springer Theses. ISBN 978-3-319-09539-4. Berlin: Springer-Verlag, 2015
\bibitem[]{}  Giri K., Chakrabarti S. K.,2013,Monthly Notices of the Royal Astronomical Society, Volume 430, Issue 4, p.2836-2843
\bibitem[]{} Harten A., 1983, J. Comp. Phys., 49, 357
\bibitem[]{} Heyvaerts J., Norman C., 1989, ApJ, 347, 1055
\bibitem[]{} Königl A., 1989, ApJ, 342, 208
\bibitem[]{} Longcope D. W., Klapper I., 1997, ApJ, 488, 443
\bibitem[]{} Lovelace R. V. E., 1976, Nature, 262, 649
\bibitem[]{} Lynden-Bell D., 1978, Phys. Scr., 17, 185
\bibitem[]{} Molteni D., Ryu D., Chakrabarti S. K., 1996, ApJ, 470, 460
\bibitem[]{} Moreno-Insertis F., Schuessler M., Ferriz-Mas A., 1992, A\&A,
264, 686
\bibitem[]{} Naik S., Rao A. R., 2000, A\&A, 362, 691
\bibitem[]{} Nandi A., Chakrabarti S. K., Vadawale S. V., Rao A. R., 2001,A\&A, 380, 245
\bibitem[]{} Paczy\'nsky B., Wiita P. J., 1980, A\&A, 88, 23
\bibitem[]{} Parker E. N., 1955, ApJ, 121, 491
\bibitem[]{} Rees M. J., Begelman M. C., Blandford R. D., Phinney E. S.,
1982, Nature, 295, 17
\bibitem[]{} Ryu D., Brown G. L., Ostriker J. P., Loeb A., 1995, ApJ, 452,
364
\bibitem[]{} Ryu D., Chakrabarti S. K., Molteni D., 1997, ApJ, 474, 378
\bibitem[]{} Sakimoto P. J., Coroniti F. V., 1989, ApJ, 342, 49
\bibitem[]{} Shibata K., Uchida Y., 1985, PASJ, 37, 31
\bibitem[]{} Shibata K., Uchida Y., 1986, PASJ, 38, 631
\bibitem[]{} Shibata K., Tajima T., Matsumoto R., 1990, ApJ, 350, 295
\bibitem[]{} Vadawale S. V., Rao A. R., Nandi A., Chakrabarti S. K., 2001,
A\&A, 370,  L17
\bibitem[]{} Vadawale S. V., Rao A. R., Naik S., Yadav J. S., Ishwara-Chandra
C. H., Pramesh Rao A., Pooley G. G., 2003, ApJ, 597, 1023
\bibitem[]{} You S., Yun G. S., Bellan P. M., 2005, Physical Review Letters,
95, 045002
\end{thebibliography}




%



\bsp	
\label{lastpage}
\end{document}